\documentclass[aps,pra,longbibliography,twocolumn,floatfix,amsmath,superscriptaddress,amssymb,bibnotes]{revtex4-1}

\usepackage{graphicx}
\usepackage{siunitx}
\usepackage{xspace}
\usepackage{times}
\usepackage[utf8]{inputenc}
\usepackage{color}
\usepackage{color}
\usepackage{bm}

\usepackage[colorlinks=true,linkcolor=black,citecolor=blue,urlcolor=blue,pdfborder={0 0 0}]{hyperref}

\newcommand{\Tr}{\text{Tr}}

\newcommand{\rs}{\rm \scriptscriptstyle}

\newcommand{\ket}[1]{\ensuremath{\left| #1\right\rangle}\xspace}       
\newcommand{\ketbra}[2]{\left|#1\middle\rangle\middle\langle #2\right|} 

\newcommand{\vc}[1]{\mathbf{#1}} 

\newcommand{\Eref}[1]{Eq.\,(\ref{#1})}
\newcommand{\appref}[1]{Appendix\,\ref{#1}}
\newcommand{\figref}[1]{Fig.\,\ref{#1}}

\newcommand{\itpthree}{Institut f\"ur Theoretische Physik III and Center for Integrated Quantum Science and Technology, Universität Stuttgart, 70569 Stuttgart, Germany}
\newcommand{\pifive}{5. Physikalisches Institut and Center for Integrated Quantum Science and Technology, Universität Stuttgart, 70569 Stuttgart, Germany}
\newcommand{\sdu}{Department of Physics, Chemistry and Pharmacy, University of Southern Denmark, 5230 Odense, Denmark}

\begin{document}

\title{Free-Space Quantum Electrodynamics with a single Rydberg superatom}

\author{Asaf Paris-Mandoki}
\thanks{These authors contributed equally to this work.}
\affiliation{\pifive}
\affiliation{\sdu}
\author{Christoph Braun}
\thanks{These authors contributed equally to this work.}
\affiliation{\pifive}
\affiliation{\sdu}
\author{Jan Kumlin}
\affiliation{\itpthree}
\author{Christoph Tresp}
\affiliation{\pifive}
\affiliation{\sdu}
\author{Ivan Mirgorodskiy}
\affiliation{\pifive}
\author{Florian Christaller}
\affiliation{\pifive}
\affiliation{\sdu}
\author{Hans Peter B\"uchler}
\affiliation{\itpthree}
\author{Sebastian Hofferberth}
\email[]{hofferberth@sdu.dk}
\affiliation{\pifive}
\affiliation{\sdu}
\date{\today}

\begin{abstract}
The interaction of a single photon with an individual two-level system is the textbook example of quantum electrodynamics. Achieving strong coupling in this system so far required confinement of the light field inside resonators or waveguides. Here, we demonstrate strong coherent coupling between a single Rydberg superatom, consisting of thousands of atoms behaving as a single two-level system due to the Rydberg blockade, and a propagating light pulse containing only a few photons. The strong light-matter coupling in combination with the direct access to the outgoing field allows us to observe for the first time the effect of the interactions on the driving field at the single photon level. We find that all our results are in quantitative agreement with the predictions of the theory of a single two-level system strongly coupled to a single quantized propagating light mode.
\end{abstract}

\maketitle

The interaction between a single emitter and individual photons is a fundamental process in nature \cite{Cohen-Tannoudji1997}, underlying many phenomena such as vision and photosynthesis as well as applications including imaging, spectroscopy or optical information processing and communication. In the strong coupling limit, where the coherent interaction between a single photon and an individual emitter exceeds all decoherence and loss rates, a single emitter can function as interface between stationary and flying qubits, a central building block for future quantum networks \cite{Kimble2008c,Rempe2015b}. Such a quantum optical node is able to mediate effective photon-photon interactions, thus enabling deterministic all-optical quantum gates \cite{Ritter2016,Dayan2014,Lukin2014}.

One groundbreaking scheme to achieve strong coupling is the use of electromagnetic (EM) cavities, where the photons are trapped within the finite volume of a high-finesse resonator. The physics of these systems is captured by the seminal Jaynes-Cummings model \cite{Shore1993}, which has been experimentally realized and extensively studied in atomic cavity quantum electrodynamics (QED) \cite{Raimond2001} and more recently in circuit QED systems combining on-chip microwave resonators with superconducting two-level systems \cite{Wallraff2004a,Schoelkopf2008b}. Achieving a strong interaction between a propagating photon and a single emitter opens the possibility to realize novel quantum-optical devices where atoms process photonic qubits on the fly and facilitate the preparation of non-classical states of light \cite{Zoller2015}. However, mode matching between the input field and the dipolar emission pattern of the quantum emitter in free space is challenging and has so far limited the achievable coupling strength \cite{Unlu2007,Kurtsiefer2008,Sandoghdar2016}. Waveguide QED systems seek to overcome this limitation by transversely confining the propagating EM mode coupled to one or more emitters \cite{Wallraff2013b,Rauschenbeutel2014,Sandoghdar2014,Lodahl2015c,Kimble2015b,Lukin2016d,Lukin2016}.

Here we report on the realization of coherent coupling between a propagating few-photon optical field and a single Rydberg superatom in free space. By exploiting the Rydberg blockade effect in an atomic ensemble \cite{Lukin2001c,Saffman2009,Grangier2009,Kuzmich2012c}, which allows at most a single excitation shared among all $N$ constituents, we turn $ \sim 10^4$ individual ultracold atoms into a single effective two-level quantum system. The collective nature of this excitation enhances the coupling of the light field to the superatom by a factor of $\sqrt{N}$ compared to the single-atom coupling strength and guarantees an enhanced directed emission in the forward direction \cite{Lukin2001c,Wodkiewicz2006}. The resulting large coupling enables us to drive Rabi oscillations of the single superatom with a few-photon probe pulse and to observe for the first time the effects of the coherent emitter-photon interaction on the photon-photon correlations of the outgoing field. We show that our system is well described by the theory of a single quantum emitter strongly coupled to a one-dimensional quantized light mode and that the light-matter coupling we achieve in free space is competitive to state-of-the-art optical waveguide QED systems \cite{Lodahl2015c,Lukin2016d,Lukin2016}.

\begin{figure}
\centering
\includegraphics[width=\columnwidth]{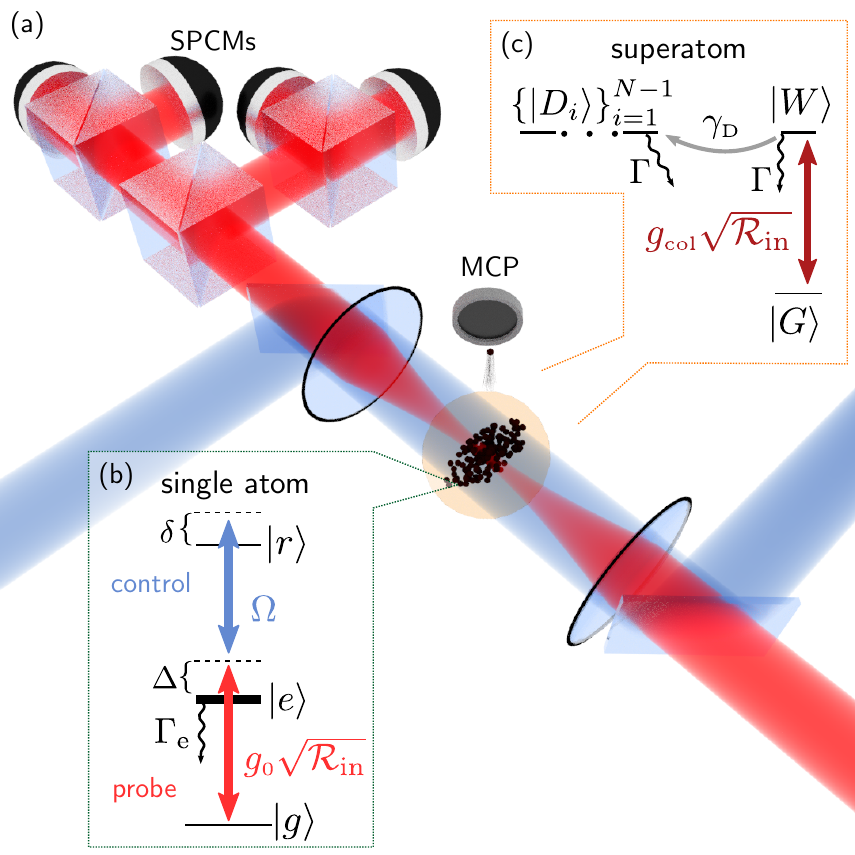}
\caption{Experimental setup and level scheme. (a) An ensemble of laser-cooled atoms is confined within a blockade volume using an optical dipole trap. Single-photon counting modules (SPCMs) are used to detect the light that interacts with the atoms while a multi-channel plate (MCP) detects the Rydberg atoms after ionization. (b) A few-photon probe field (red) and a strong control field (blue) couple the single atom ground state $\ket{g}$ to a Rydberg state $\ket{r}$. Their respective Rabi frequencies are $g_{\rs 0}\sqrt{\mathcal{R}_\mathrm{in}}$ and $\Omega$, where $g_{\rs 0}$ is the single atom coupling constant for the probe field and $\mathcal{R}_\mathrm{in}$ is the photon rate. (c) Due to the Rydberg blockade, the whole ensemble collectively behaves like a two-level system with many-body states $\ket{G}$ and $\ket{W}$ including a loss channel to a set of collective dark states $\left\{\ket{D_i}\right\}_{i=1}^{N-1}$.\label{Fig1:Setup}}
\end{figure}

We implement our single Rydberg superatom by focusing a weak \SI{780}{\nano\meter} probe field (beam waist $w_\mathrm{probe}=\SI{6.5}{\micro\meter}$), together with a strong counter-propagating control field at \SI{480}{\nano\meter}  (beam waist $w_\mathrm{control}=\SI{14}{\micro\meter}$) into an optically trapped ensemble of ultracold $^{87}$Rb atoms ($T=\SI{6}{\micro\kelvin}$) (\figref{Fig1:Setup}a and \appref{app:prerparation}). The few-photon coherent probe field, with a photon rate $\mathcal{R}_\mathrm{in}$, couples the ground $\left|g\right\rangle=\left|5S_{1/2},F=2,m_F=2\right\rangle$  and intermediate $\left|e\right\rangle=\left|5P_{3/2},F=3,m_F=3\right\rangle$ states with a Rabi frequency $g_{\rs 0}\sqrt{\mathcal{R}_\mathrm{in}}$, where $g_{\rs 0}$ is the single-atom--single-photon coupling constant, determined by the geometry of the setup. The control field provides coupling between $\ket{e}$ and the Rydberg state $\left|r\right\rangle=\left|111S_{1/2},m_J=1/2\right\rangle$ with Rabi frequency $\Omega=2\pi\times\SI{10}{\mega\hertz}$ (\figref{Fig1:Setup}b). Using a large intermediate-state detuning $\Delta=2\pi\times\SI{100}{\mega\hertz}\gg\Gamma_e,\Omega$, the intermediate state can be adiabatically eliminated. Setting the two-photon detuning $\delta$ to Raman resonance the dynamics for each atom simplifies to those of a resonantly coupled two-level system between $\ket{g}$ and $\ket{r}$ with effective Rabi frequency $ g_{\rs 0}\sqrt{\mathcal{R}_\mathrm{in}}\Omega/(2\Delta)$. The decay of $\ket{r}$ is dominated by spontaneous Raman decay via the $\ket{e}$ level with rate $\Gamma = \Omega^2/(2\Delta)^2 \Gamma_\mathrm{e}$.

The interaction between Rydberg atoms results in a blockade volume inside which only a single excitation is allowed \cite{Lukin2001c,Saffman2009,Grangier2009,Kuzmich2012c}. In our setup, both the transverse probe beam diameter and the longitudinal extent of the atomic cloud are smaller than the radius of the blockade volume, collectively coupling $N \approx 10^4$ atoms within this volume to the propagating light mode. Specifically, the $N$-body ground state $\ket{G}=\ket{g_1,\dots, g_N}$  couples only to one many-body excited state $\ket{W} = \frac{1}{\sqrt{N}}\sum_{j=1}^N e^{i \vc{k}\cdot\vc{x}_j}\ket{j}$, where $\ket{j}=\ket{g_1,\dots,r_j,\dots,g_N}$ is the state with the $j$-th atom in $\ket{r}$ and all others in $\ket{g}$, $\vc{k}$ is the sum of the wavevectors of the probe and control fields and $\vc{x}_j$ denotes the position of the $j$-th atom. Ultimately, the ensemble of $N$ atoms acts as a single two-level superatom coupled to the probe light via the collective coupling constant $g_{\rs col}=\sqrt{N} g_{\rs 0} \Omega/(2\Delta)$ (\figref{Fig1:Setup}c). In addition to $\ket{G}$  and $\ket{W}$, the Hilbert space describing the system contains $N-1$ collective dark states $\left\{\ket{D_i}\right\}_{i=1}^{N-1}$ formed by linear combinations of $\left\{\left|j\right\rangle\right\}_{j=1}^{N}$. While these states still contain an excitation that blocks the medium, they are not coupled to the probe light. The exchange of virtual photons between atoms has been shown to provide coupling between the $\ket{W}$ state and the collective dark states \cite{Lehmberg1970,Scully2008,Cirac2008,Molmer2013}, which can alter the decay rate of the bright state \cite{Guerin2016,Havey2016}. Additionally, inhomogeneous dephasing acting on individual atoms can irreversibly drive the ensemble from $\ket{W}$ into the manifold of dark states $\left\{\ket{D_i}\right\}_{i=1}^{N-1}$, which enables the system to function as a single-photon absorber \cite{Honer2011,Tresp2016a}.

\begin{figure}
\centering
\includegraphics[width=\columnwidth]{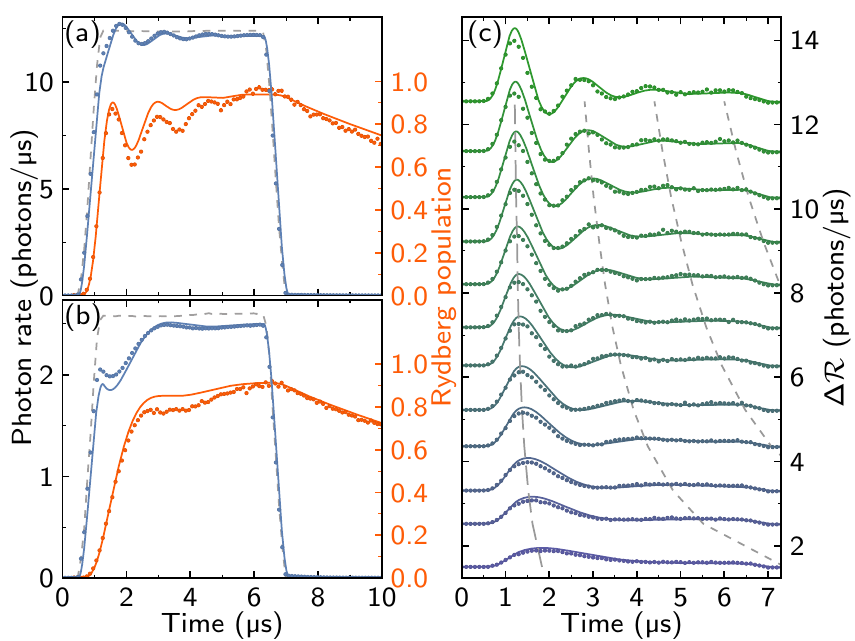}
\caption{Time evolution of photon signal and Rydberg population. (a and b) Time traces of the outgoing probe photon rate (blue points) and Rydberg population (orange points) for input pulses (dashed gray line) with peak photon rates $\mathcal{R}_\mathrm{in}= \SI{12.4}{\per\micro\second}$ (a) and $\mathcal{R}_\mathrm{in}=\SI{2.6}{\per\micro\second}$ (b), corresponding to mean number $\bar{N}_\mathrm{ph}$ of 71.6 and 15.1  photons in the pulse. The Rabi oscillation of the single superatom is visible both in the excited state population and in the modulation of the transmitted probe light. Solid lines are fits to the data using our model. (c) Difference signal $\Delta \mathcal{R}(t)$ between the incoming and outgoing pulses (dots) for different input photon rates $\mathcal{R}_\mathrm{in}$. Each data set is vertically shifted by the corresponding $\mathcal{R}_\mathrm{in}$. Solid lines are again the result of fitting the full data set with our theory model using a single set of fit parameters. Dashed lines indicate the expected positions of the Rabi oscillation peaks based on the fitted parameters, showing the scaling of the Rabi period with $1/\sqrt{\bar N_\mathrm{ph}}$. Error bars in (a-c) are SEM and are smaller than the data points.\label{Fig2:RabiOsc}}
\end{figure}

To observe the coherent dynamics of the superatom, we send a Tukey-shaped probe pulse with a peak photon rate $\mathcal{R}_\mathrm{in}$ into the atomic cloud. After its interaction with the ensemble, the probe light is collected by four single-photon counters (\figref{Fig1:Setup}a). Alternatively, the Rydberg population in the ensemble at any time is measured by counting the ions produced by a fast field-ionization pulse with a micro-channel plate detector (see \appref{app:pulsedexp}). In \figref{Fig2:RabiOsc}a,b we show average photon time traces and ion signals for $\mathcal{R}_\mathrm{in}= \SI{12.4}{\per\micro\second}$ and $\mathcal{R}_\mathrm{in}=\SI{2.6}{\per\micro\second}$. First, we observe the collectively enhanced Rabi oscillation of the Rydberg population \cite{Saffman2009,Grangier2009,Kuzmich2012c,Ott2015,Gross2015c,Browaeys2016b}. Importantly, the number of Rydberg atoms throughout the whole pulse stays below one, showing that our ensemble is indeed fully blockaded and well-described as a single superatom. The coherent dynamics of the system also cause a periodic modulation of the outgoing photon rate $\mathcal{R}_\mathrm{out}$. \figref{Fig2:RabiOsc}c shows this modulation $\Delta \mathcal{R}(t) = \mathcal{R}_\mathrm{in}(t)-\mathcal{R}_\mathrm{out}(t)$ for a range of input rates, down to $\mathcal{R}_\mathrm{in}=\SI{1.5}{\micro\second}^{-1}$, which corresponds to a mean number of photons $\bar{N}_\mathrm{ph}=9$ in the probe pulse.

In order to quantitatively describe our results, we consider the Hamiltonian of a single two-level system coupled to a quantized light field
\begin{equation}
H = \int \frac{dk}{2\pi} \hbar c k a_k^\dagger a_k+\frac{\hbar g_{\rs col}}{2}\left(E^\dagger(0)\sigma_{\rs GW}+E(0)\sigma_{\rs GW}^\dagger\right),
\label{eq:hamiltonian}
\end{equation}
where $a_k$ and $a_k^{\dagger}$ are photon annihilation and creation operators, $E(x)=\frac{\sqrt{c}}{2\pi} \int e^{ikx} a_k \, dk $ is the electric field operator measured in $\sqrt{\mathrm{photons}/\mathrm{time}}$ and $\sigma_{\alpha\beta}=\ketbra{\alpha}{\beta}$. Since the probe photons irreversibly leave after a single pass through the system, we can solve and trace out the time-dependence of the photonic part (see \appref{app:derivation_of_mastereq}). For a coherent input state, we obtain a master equation for the atomic density matrix \cite{Zoller2015,Cirac2015}
\begin{align}
\begin{split}
\partial_t \rho(t) = &-\frac{i}{\hbar}\left[H_0(t),\rho(t)\right]+(\kappa+\Gamma) \mathcal{L}\left[\sigma_{\rs GW}\right]\rho(t)
 \\ &+ \gamma_{\rs D} \mathcal{L}\left[\sigma_{\rs DW}\right]\rho(t)
+ \Gamma\mathcal{L}\left[\sigma_{\rs GD}\right]\rho(t),
\end{split}\label{eq:master}
\end{align}
where $\mathcal{L}[\sigma]\rho = \sigma\rho\sigma^\dagger-(\sigma^\dagger\sigma\rho+\rho\sigma^\dagger\sigma)/2$ is the Lindblad superoperator and the effective Hamiltonian is
\begin{equation}
H_0(t) = \hbar \sqrt{\kappa}\left(\alpha^*(t)\sigma_{\rs GW}+\alpha(t)\sigma_{\rs GW}^\dagger\right),
\label{eq:coherent_hamiltonian}
\end{equation}
with the coherent field amplitude $\alpha(t)$ related to the time-dependent mean photon rate by $|\alpha(t)|^2=\mathcal{R}_ {in}(t)$ and the rate of emission $\kappa = g_{\rs col}^2/4$ of the two-level system into the strongly coupled mode. In addition to this intrinsic decay channel, which is derived from the Hamiltonian in \Eref{eq:hamiltonian}, we phenomenologically add the spontaneous Rydberg atom decay rate $\Gamma$ of the excited state and the dephasing rate $\gamma_{\rs D}$ of the superatom state $\ket{W}$ into the manifold of dark states $\ket{D}$. The Rydberg population is then given by $\rho_{\rs WW}+\rho_{\rs DD}$, while the outgoing electric field is
\begin{equation}
E(t) = \alpha(t)-i\sqrt{\kappa}\sigma_{\rs GW}(t).
\label{eq:el_field}
\end{equation}
The equal-time expectation values for the electric field operator therefore reduce to the determination of equal-time correlations in
$\sigma_{\rs{GW}}^{\dag}(t)$, which are obtained by the numerical solution of \Eref{eq:master}. In particular, the expectation
value of the photon flux at retarded time $s = t - x/c$ is given by
\begin{equation}
\begin{split}
\langle E^\dagger(s) E(s) \rangle =\vert \alpha(s) \vert^2 + \kappa \, \langle  \sigma_{\rs{GW}}^{\dag}(s)  \sigma_{\rs{GW}}(s) \rangle\\ - i \sqrt{\kappa}   \left[ \alpha^{*}(s) \langle \sigma_{\rs{GW}}(s) \rangle -   \alpha(s) \langle  \sigma^{\dag}_{\rs{GW}}(s) \rangle \right]   .
\label{eq:intensity}
\end{split}
\end{equation}
The solid lines in \figref{Fig2:RabiOsc}a,b,c, are the result of fitting the above model to the respective sets of time traces. For the set in \figref{Fig2:RabiOsc}a,b, we obtain the values $\kappa=\SI{0.428}{\per\micro\second}$, $\Gamma=\SI{0.069}{\per\micro\second}$, and $\gamma_{\rs D}=\SI{1.397}{\per\micro\second}$.


\begin{figure}
\centering
\includegraphics[width=\columnwidth]{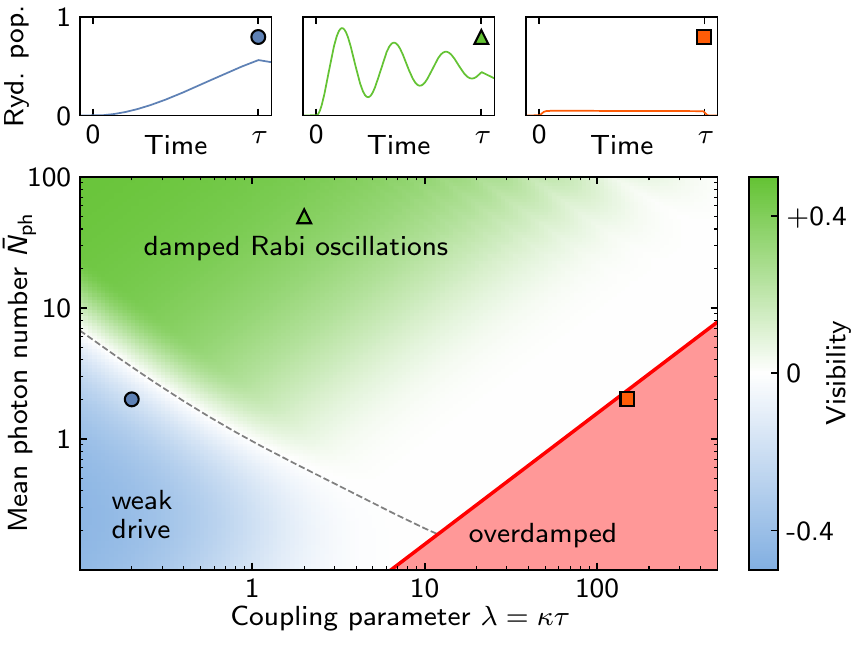}
\caption{Dynamical phase diagram of a driven atom in free space. (bottom) The diagram shows the visibility of Rabi oscillations, defined as $\max_{0\leq t \leq\tau}\left[\rho_{\rs WW}(t)\right]-\rho_{\rs WW}(t=\infty)$, of an ideal ($\Gamma=\gamma_{\rs D}=0$) atom driven by a propagating field. In contrast to cavity QED, the coupling and the decay of photons from the system are not independent in free-space and waveguide QED. For large coupling to the propagating mode ($\lambda = \kappa \tau \gg 1$) the enhanced emission into this mode results in an overdamped system, where the number of photons required to observe Rabi oscillations increases with coupling strength. For $\lambda \ll 1$, a large number of photons is required to drive the system with a $\pi$-pulse, defining a crossover (dashed line) between the regime of damped Rabi oscillations and the weak driving regime at lower mean photon number. For our experiment, we find $\lambda = 2.2$. (top) Examples of the variation of the Rydberg population with time for the points indicated in the main diagram.
\label{Fig3:Diagram}}
\end{figure}

For a single atom in free space the coupling with a photon, quantified by $\kappa$, can maximally become as large as the spontaneous decay rate of the bare atom $\Gamma$ in the case of perfect mode matching \cite{Leuchs2013}. In the superatom case, the coupling $g_{\rs col} \sim \sqrt{\kappa}$ and the decay $\kappa$ into a specific mode can be boosted solely through the collective enhancement of the atom-light interaction, without any confinement of the propagating light. As a consequence, the superatom spontaneously emits with probability $\beta = \kappa/(\kappa+\Gamma)=0.86$ into the forward direction of the strongly coupled mode, while loss of photons due to scattering out of the propagating mode with rate $\Gamma$ is minimal (see \appref{app:gaussian_enhancement}). The main decoherence source in our current implementation is the superatom dephasing $\gamma_{\rs D}$, we expect that thermal motion of the individual atoms in the superatom are the leading mechanism for this dephasing, which could  be significantly reduced by technical improvement of our setup \cite{Bloch2009,Kuzmich2013b}. However, a coherent virtual exchange of photons can provide an additional coherent dynamics for the superatom \cite{Lehmberg1970,Scully2008,Cirac2008,Molmer2013}. Given the excellent agreement between the experimental data and our model, we conclude that in the present experimental regime the potential influence of this coherent term is well accounted for by the phenomenological dephasing rate $\gamma_D$.

In contrast to cavity QED, in free-space and waveguide QED, an increase of the coupling $g_{\rs col}$ necessarily increases the decay rate $\kappa$, resulting in an intrinsic damping of these systems preventing perfect transfer of a photonic qubit to a matter qubit within a finite time \cite{Scarani2011,Leuchs2013,Kurtsiefer2016}. To further illustrate this point, \figref{Fig3:Diagram} shows the visibility of Rabi oscillations of an ideal ($\Gamma=\gamma_{\rs D}=0$) two-level atom (see \appref{app:analytic_sol}) as a function of the dimensionless coupling parameter $\lambda = \kappa \tau$,  where $\tau$ is the length of the incoming pulse, and the mean photon number in the pulse $\bar{N}_\mathrm{ph} = \mathcal{R}_\mathrm{in} \tau$. For $\lambda \gg 1$ the atom decays very quickly and photons exhibit correlations only on a timescale $1/\kappa$. This results in an \textit{overdamped} regime, where the system settles to a nonzero probability to find the superatom in the excited state without undergoing any Rabi oscillations. In the opposite limit $\lambda \ll 1$, a minimum number of photons is required to drive a $\pi$-pulse, which defines a crossover (dashed line in \figref{Fig3:Diagram}) between the regime with Rabi oscillations and the \textit{weak driving} regime, where the excitation probability remains below its steady state value during the pulse duration $\tau$. For our current experiments $\lambda = 2.2$ (see \appref{app:data_and_statistics}), thus placing our setup close to the ideal regime where multiple Rabi cycles are observable within the decay time $1/\kappa$ for very low $\bar{N}_\mathrm{ph}$.


\begin{figure*}
\includegraphics[width=\textwidth]{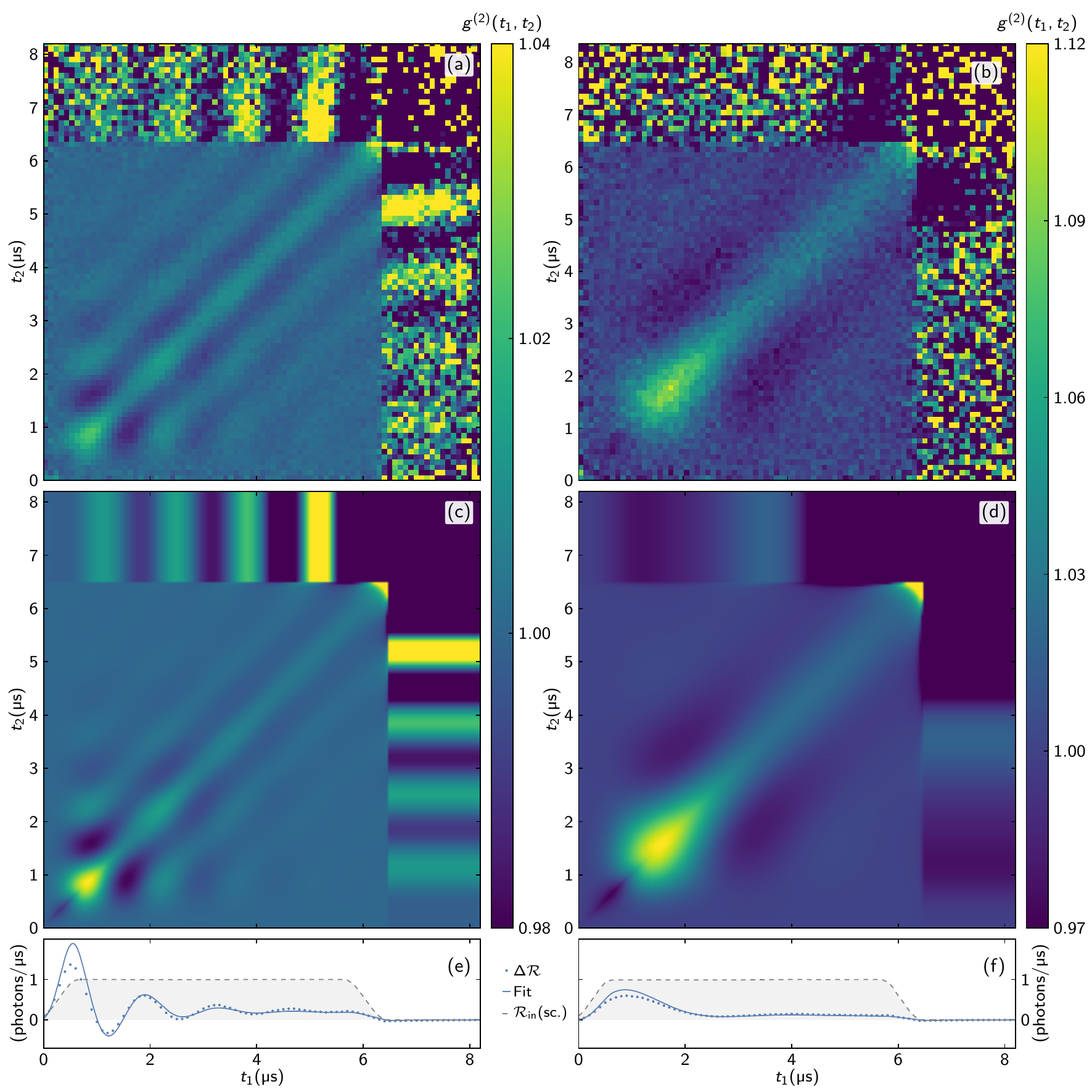}
\caption{Correlations of the outgoing probe field. (a and b) Measured two-time correlations $g^{(2)}(t_1,t_2)$ for pulses with $\mathcal{R}_\mathrm{in}= \SI{12.4}{\per\micro\second}$ and $\mathcal{R}_\mathrm{in}=\SI{2.5}{\per\micro\second}$ corresponding to the time traces in \figref{Fig2:RabiOsc}a,b. (c and d) Corresponding calculated correlations functions using the values of $\kappa$, $\Gamma$ and $\gamma_{\rs D}$ obtained by fitting the time traces in \figref{Fig2:RabiOsc}a,b. (e and f) Measured (dotted) and simulated (solid) $\Delta \mathcal{R}_\mathrm{in}$ together with scaled input pulses (dashed gray) for reference. \label{Fig4:g2}}
\end{figure*}

Since we have access to the full counting statistics of the outgoing light, we can investigate how the dynamics of the coupled system results in correlations between emerging probe photons. We show in \figref{Fig4:g2}a,b the measured two-time correlation functions $g^{(2)}(t_1,t_2)$ for $\mathcal{R}_\mathrm{in}= \SI{12.4}{\per\micro\second}$ and $\mathcal{R}_\mathrm{in}=\SI{2.5}{\per\micro\second}$. The periodic structure of bunching and antibunching can be understood as the rearrangement of photons in the coherent input beam due to the absorption and stimulated emission by the superatom, emphasizing the long coherence time of the superatom-photon interaction. \figref{Fig4:g2}c,d show the corresponding calculated correlations from our model. The evaluation of the two-point correlation function, in contrast to the intensity expectation value, requires the determination of correlations of the operators $\sigma_{\rs GW}^{\dag}(t)$ at different times, i.e.
\begin{equation}
  g^{(2)}(s_1,s_2)= \frac{\langle E^{\dag}(s_1)E^{\dag}(s_2) E(s_2) E(s_1) \rangle}{\langle E^\dagger(s_1) E(s_1) \rangle \langle E^\dagger(s_2) E(s_2) \rangle}.
\end{equation}
It is a remarkable property of a single atom coupled to a single photonic mode, that these expectation values can be determined
by the quantum regression theorem without involving any additional approximations. While in general the quantum regression theorem
relies on a Born approximation quenching the correlations between the bath and the system \cite{Lax1968}, here, the emitted photons never interact with the superatom again, which is exactly the requirement for the validity of the quantum regression theorem \cite{Cirac2015}. To verify this statement we use the exact diagonalization of the Hamiltonian in \Eref{eq:hamiltonian} by means of the Bethe ansatz \cite{Rupasov1984,Yudson1985,Yudson2008} to obtain the wave function of the outgoing pulse for few-photon Fock states. We then find perfect agreement for the correlation function derived from the exact wave function via the Bethe ansatz and the above derivation using the quantum regression theorem for a coherent state with low mean photon number. Furthermore, the theory results are in excellent agreement with the experimental data, including the correlations beyond the duration of the pulse. These originate from collective spontaneous emission of single photons after the input pulse has left the sample, which can only occur if the superatom is in state $\ket{W}$ at the end of the driving pulse. The observed correlations indicate that, due to the effective photon-photon interaction mediated by the single superatom, photons separated by up to $\SI{5}{\micro\second}$ in time become entangled \cite{Fan2012}. To illustrate this point, consider two incoming photons: the first photon passing the atom results in a superposition state of the photon either being absorbed and the superatom excited or the photon having passed the atom without exciting it \cite{Scully2008}. The second photon passing by then has restricted options depending whether the superatom state is already occupied or not, i.e. it can only be absorbed if the first photon was not, resulting in spatial entanglement between the two photons, mediated by their subsequent interaction with a single two-level system.


In conclusion, our measurements demonstrate the realization of strong light-matter coupling in free-space, through collective enhancement of the coupling strength wihtout any confining structures for the propagating light mode. The tunable dephasing of the superatom into dark states creates additional functionality beyond the conventional two-level system \cite{Honer2011,Tresp2016a}. The scaling of our system to complex arrangements of multiple superatoms is straightforward and paves the way towards quantum optical networks \cite{Kimble2008c,Rempe2015b} and the realization of strongly correlated states of light and matter \cite{Lukin2014}. The directionality of the superatom emission can be used to implement a cascaded quantum system for dissipative entanglement generation among the superatoms \cite{Zoller2012,Zoller2017}.

\begin{acknowledgments} We thank Ofer Firstenberg and J\"urgen Eschner for stimulating discussions and Hannes Gorniaczyk for contributions to the experiment. This work is funded by the German Research Foundation (Emmy-Noether-grant HO 4787/1-1, GiRyd project HO 4787/1-3, SFB/TRR21 project C12), the Ministry of Science, Research and the Arts of Baden-W\"{u}rttemberg (RiSC grant 33-7533.-30-10/37/1), European Union under the ERC consolidator grant SIRPOL (grant N. 681208), and in part by the National Science Foundation under Grant No. NSF PHY-1125915.
\end{acknowledgments}

\appendix
\section{Preparation of a single superatom}\label{app:prerparation}
To prepare the ultracold atomic ensemble that forms our single Rydberg superatom, we initially load $^{87}$Rb atoms into a magneto-optical trap (MOT)  from $\SI{E-10}{\milli\bar}$ rubidium background pressure in an ultra-high vacuum chamber, resulting in $5\times 10^6$ laser-cooled atoms after $\SI{1}{\second}$ of loading. After compressing the MOT by increasing the gradient of the quadrupole magnetic field, the atoms are loaded into a dipole trap formed by two crossed \SI{1070}{\nano\meter} beams, intersecting under an angle of $\SI{31.4}{\degree}$, and an additional elliptic dimple beam at $\SI{855}{\nano\meter}$ perpendicular to the long axis of the original trap. Subsequently, atoms are further cooled by forced evaporation for $\SI{700}{\milli\second}$ by reducing the power of the two \SI{1070}{\nano\meter} dipole trap beams. During evaporative cooling the cloud is additionally cooled during two stages of Raman Sideband Cooling. The stages are $\SI{10}{\milli\second}$ long and occur after $\SI{89}{\milli\second}$ and $\SI{589}{\milli\second}$ of evaporation, eventually reaching a final cloud temperature of $\SI{6}{\micro\kelvin}$. The final atomic cloud contains 25000 atoms in a pancake shaped harmonic trap, with a Gaussian density profile with widths $\sigma_z=\SI{6}{\micro\metre}$ and $\sigma_r=\SI{10}{\micro\metre}$. The Rydberg state $111S_{1/2}$ was chosen such that the longitudinal diameter of the cloud as well as the transverse diameter of the probe beam are significantly smaller than the minimum Rydberg blockade radius $r_b=\SI{25.5}{\micro\meter}$ at highest photon rate $\mathcal{R}_\mathrm{in}=\SI{12.4}{\per\micro\second}$, resulting in a fully blockaded ensemble of atoms coupled to the probe (waist $w_\mathrm{probe}=\SI{6.5}{\micro\meter}$) and control light (waist $w_\mathrm{control}=\SI{14}{\micro\meter}$).

\section{Pulsed few-photon experiments}\label{app:pulsedexp}
After preparation of the atomic sample, 1000 individual experiments as described in the main text are performed within a time of $\SI{100}{\milli\second}$. For each experiment, the optical trapping beams are turned off for $\SI{14}{\micro\second}$, with the atoms being recaptured after each pulse sequence. The Tukey-shaped probe pulses have rise and fall times of $\SI{0.8}{\micro\second}$ and an uptime of $\SI{5}{\micro\second}$. After $\SI{2}{\micro\second}$ initial wait time to fully turn off the crossed dipole trap, the control light is turned on $\SI{2}{\micro\second}$ before the probe pulse and remains on during the full remaining $\SI{12}{\micro\second}$. Alternatively, for measuring the Rydberg population a field ionization pulse is applied at a varying time and the produced ions are detected on a multi-channel plate (MCP). An electric field amplitude  $F > \SI{85}{\volt\per\cm}$ ensures that all Rydberg atoms are ionized. After 1000 individual experiments the atomic cloud is released and after $\SI{10}{\milli\second}$ wait time we again perform 1000 experiments without an atomic cloud for reference. Due to the finite detuning from the intermediate state, the probe transmission in the absence of the control field is $\SI{99}{\percent}$. We detect a photon after it passes through the atomic ensemble with an overall detection efficiency of $\SI{29}{\percent}$, after field ionizing an individual Rydberg atom is detected by the MCP with an overall detection efficiency of $\SI{30}{\percent}$.

\section{Data analysis and statistics}\label{app:data_and_statistics}
 Each datapoint of the photon time traces in \figref{Fig2:RabiOsc}a,b corresponds to an average of $36.85\times10^6$ and $21.73\times 10^6$ individual measurements respectively. From these datasets we also obtain the correlation functions in \figref{Fig4:g2}a,b. Each data point of the Rydberg population measurements in \figref{Fig2:RabiOsc}a,b represents an average over $32\times 10^3$ and $61\times 10^3$ experiments. As stated in the main text, we fit our numerical model to the experimental data and obtain a common set of parameters for \figref{Fig2:RabiOsc}a,b, namely $\kappa=\SI{0.428}{\per\micro\second}$, $\Gamma=\SI{0.069}{\per\micro\second}$, and $\gamma_{\rs D}=\SI{1.397}{\per\micro\second}$. These parameters are used for the calculated correlation functions in \figref{Fig4:g2}a,b.
 Each data point in the photon traces in \figref{Fig2:RabiOsc}c represents the average of $1.86\times 10^6$ experiments. Here, we fit with a single set of parameters to all shown traces simultaneously, obtaining $\kappa=\SI{0.322}{\per\micro\second}$, $\Gamma=\SI{0.069}{\per\micro\second}$, and $\gamma_{\rs D}=\SI{1.326}{\per\micro\second}$. The differences in $\kappa$ and $\gamma_{\rs D}$ between the two data sets stem from slightly different number of atoms $N$ in the superatom for the two experiment runs. The single-atom decay $\Gamma$ is set solely by the control laser parameters and Rydberg lifetime and thus does not change between runs. For all shown data the standard error of the mean (SEM) is smaller than the size of the displayed dots.
 For the calculation of the dimensionless coupling parameter $\lambda = \kappa \tau$ we use the mean of the two measured $\kappa$ and an effective length of the probe pulse of $\tau=\SI{5.8}{\micro\second}$.

\section{Derivation of master equation}\label{app:derivation_of_mastereq}
We start with the Hamiltonian \Eref{eq:hamiltonian}, which describes the coherent coupling of the superatom to the optical mode of the incoming laser field. The derivation of the master equation closely follows the methods described in standard textbooks \cite{Gardiner2004}, and recent publications on atom-light coupling in one-dimensional wave guides \cite{Zoller2015,Cirac2015}. The first step is to derive the Heisenberg equation of motion for the photonic field operators
\begin{equation}
\partial_t a_k(t) = - \frac{i}{\hbar} \left[a_k, H \right] = -i c k a_k(t) - i \sqrt{\kappa c}\,  \sigma_{\rs{GW}}(t)
\end{equation}
with $\sigma_{\rs{GW}} = |G\rangle\langle W|$ and the coupling strength $\sqrt{\kappa}= g_{\rs{col}}/2$. This equation has a simple solution, which leads to a connection between the outgoing electric field and the operator $\sigma_{\rs{GW}}$, which describes the coherences in the superatom,
\begin{equation}
\begin{split}
a_k(t) &=  e^{-i c k (t-t_0)} a_k(t_0) \\&- i \sqrt{\kappa c} e^{-i c k (t-t_0)} \int_{t_0}^t ds \, e^{i c k (s-t_0)} \sigma_{\rs{GW}}(s) .
\end{split}
\end{equation}
Here, $t_{0}$ denotes the initial time with the condition, that the incoming photon field has not yet reached the superatom.
Without loss of generality, we set $t_{0}=0$. Then, the electric field operator reduces to
\begin{widetext}
\begin{align}
\begin{split}
E(x,t) &= \bar{E}(c t-x) - i \sqrt{\kappa  }c \int_{0}^t ds \int \frac{dk}{2\pi }\, e^{- i c k (t-s) + i k x} \sigma_{\rs{GW}}(s),\\
&= \bar{E}(c t-x) - i \sqrt{\kappa} \sigma_{\rs{GW}}( t - x/c) \theta(x)\theta(ct - x) .
\label{eq:efield}
\end{split}
\end{align}
\end{widetext}
Here, $\bar{E}$ denotes the non-interacting electric field operator and $\theta(x)$ is the Heaviside function with
the definition that $\theta(0) = 1/2$.  For an incoming coherent state, the non-interacting electric field $\bar{E}(c \: t)$
can be replaced by the amplitude of the coherent field $\alpha(t) \equiv \langle  \bar{E}(c t)\rangle$, which characterizes  the incoming photon rate by $|\alpha(t)|^2=\mathcal{R}_\mathrm{in}$.
Alternatively, it would be possible to apply the well-established Mollow transformation \cite{Mollow1975} leading to the same final result for the master equation. For an arbitary operator $A$ acting on the superatom alone, its Heisenberg equation of motion reduces to
\begin{widetext}
\begin{align}
\begin{split}
\partial_t A(t) =  &- i \sqrt{\kappa} \left[ A(t), \sigma_{\rs{GW}}^{\dag}(t) \right] \alpha( t) - \frac{\kappa}{2} \left[ A(t), \sigma_{\rs{GW}}^{\dag}(t)\right]\sigma_{\rs{GW}}(t)  \\
 &- i \sqrt{\kappa} \alpha^{*}(t) \left[ A(t), \sigma_{\rs{GW}}(t) \right] + \frac{\kappa}{2} \sigma_{\rs{GW}}^{\dag}(t) \left[ A(t), \sigma_{\rs{GW}}(t)\right]\, .
\end{split}
\end{align}
\end{widetext}
The right-hand side can be split into a coherent part given by
\begin{align}
&- i \sqrt{\kappa} \left[ A(t), \sigma^{\dag}_{\rs{GW}}(t) \right] \alpha(t) - i \sqrt{\kappa} \alpha^*(t)\left[ A(t),\sigma_{\rs{GW}}(t) \right]  \notag
\\&= - \frac{i}{\hbar} \left[ A(t), H_{0}(t)\right]
\label{superatomcoupling}
\end{align}
with the Hamiltonian $H_{0}(t)$ in \Eref{eq:coherent_hamiltonian} while the remaining terms describe
the spontaneous emission  into the photonic mode.


Using the relation, $\partial_t \langle A \rangle = \Tr \left\{  A \partial_t\rho(t)\right\}$ with $\rho(t)$
the reduced density matrix for the atomic system alone, the dissipative part reduces to the well established Lindblad
form
\begin{equation}
\kappa \, \mathcal{L}[\sigma_{\rs{GW}}]  \rho(t) = \kappa \, (\sigma_{\rs{GW}} \rho(t) \sigma_{\rs{GW}}^{\dag}(t)  -
 \frac{1}{2} \{ \sigma_{\rs{GW}}^{\dag}  \sigma_{\rs{GW}}(t) , \rho(t) \} ).
\end{equation}
This term describes the enhanced spontaneous emission into the forward direction due to the collective character of the
superatom. In addition, the superatom can also decay into transverse photonic modes, which is still determined by the standard
single atom spontaneous emission rate $\Gamma$, see \appref{app:gaussian_enhancement}. Finally, we can add the dephasing into the dark states  $\left\{\ket{D_i}\right\}_{i=1}^{N-1}$ as well as the
decay by spontaneous emission of these dark states. The analysis is independent of the specific dark state the system dephases into,
and therefore we can account for the dephasing by losses into a single dark state $|D\rangle$ with a phenomenological dephasing rate $\gamma_{\rs D}$.
The microscopic mechanisms for the dephasing are on one hand doppler shifts of the atoms, as well as inhomogeneous shifts of the Rydberg state level, and residual interactions between
the individual atoms by resonant exchange interactions \cite{Lehmberg1970,Scully2008,Cirac2008,Zoller2015,Chang2015b}.

\section{Analytical solution of the master equation}\label{app:analytic_sol}

In the ideal case with $\Gamma =\gamma_{D}=0$, the master equation in \Eref{eq:master} allows for an analytical solution for a driving field $\alpha=\sqrt{\mathcal{R}_\mathrm{in}}$ switched on at time $t = 0$.  The full solution  for the probability to be in the
excited state $\rho_{\rs{WW}}$ is given by
\begin{equation}
\begin{split}
&\rho_{\rs{WW}}(t) = \frac{4 \alpha^2 \kappa  }{\kappa^2 + 8 \alpha^2 \kappa} \\
&\left( 1 - \left(\cos \Omega_{\rs{eff}} t + \frac{3 \kappa}{4 \Omega_{\rs{eff}}} \sin \Omega_{\rs{eff}} t \right) e^{-\frac{3}{4}\kappa t} \right)
\label{eq:solution_TLS}
\end{split}
\end{equation}
with the effective Rabi frequency $\Omega_{\rs{eff}} =\sqrt{4 \kappa \alpha^2  - \left(\kappa/4\right)^2} $.
Introducing the dimensionless coupling parameter $\lambda = \kappa \tau$,  where $\tau$ is the length of the incoming pulse,
and the mean photon number in the pulse $\bar{N}_\mathrm{ph} = |\alpha|^2 \tau$, the behavior of the superatom in free space can be described
by a dynamical phase diagram, \figref{Fig3:Diagram}. First, it is important to stress, that in contrast to cavity QED, the coupling and
the decay in this free space setup are not independent. This, for example,  implies that there is no `strong coupling' regime as in cavity QED, where the coupling can be increased  without affecting the spontaneous emission of the atom. Indeed, for increasing coupling strength
$\kappa$ at fixed mean photon number $\bar{N}_\mathrm{ph}$, the dissipation by the spontaneous emission increases, which reduces the visibility of the
Rabi oscillations and eventually leads to the overdamped regime at $\bar{N}_\mathrm{ph} = \lambda/64$. This overdamped regime is characterized  by an
imaginary effective Rabi frequency $\Omega_{\rs{eff}}$; the red line in \figref{Fig3:Diagram} shows this transition.
In turn for weak coupling $\lambda \ll 1$, it is required to have a large number of photons in order to drive the system with a $\pi$-pulse, which defines a
crossover between a regime with Rabi oscillations and the weak driving  regime at lower mean photon number. This crossover is illustrated by the
dashed line in \figref{Fig3:Diagram}.

In the experimentally relevant case of $\Gamma, \gamma_{D}>0$, the resulting master equation includes the additional level $|D\rangle$. This extended model can be solved to obtain the effective Rabi frequency $\Omega_{\rs{eff}} (\kappa,\Gamma, \gamma_{D})$. This solution is used to plot the lines predicting the Rabi oscillation maxima in \figref{Fig2:RabiOsc}c.

\section{Collective coupling and decay into the forward propagating mode}\label{app:gaussian_enhancement}
We start with a microscopic setup as realised in the experiment: a large number of atoms are localized within a harmonic trap. The density distribution of the ground state atoms is given by a Gaussian profile with widths $\sigma_{z}$ along the direction of the incoming light field and $\sigma_{r}$ in transverse direction with peak density $n_{0}$. Each atom is well described by a two-level atom with the ground state $|g\rangle$
and the excited Rydberg state $|r\rangle$ with the optical transition frequency $\omega = 2 \pi c/\lambda$ and wave length $\lambda$.
In the following, we describe the two states of the atoms by the field operators
$\psi^{\dag}_{g}({\mathbf r})$ for the ground state and  $\psi^{\dag}_{r}({\mathbf r})$ for the Rydberg state, respectively.
In terms of these operators, the ground state density is defined as the expectation value
\begin{equation}
 n({\mathbf r}) = \langle  \psi^{\dag}_{g}({\mathbf r}) \psi_{g}({\mathbf r})  \rangle =n_{0} e^{- z^2/2 \sigma_{z}^2}  e^{- (x^2+y^2)/2 \sigma_{r}^2}  .
\end{equation}
It is important to stress that for a thermal gas above quantum degeneracy the statistics of the operator
$\psi^{\dag}_{g}({\mathbf r})$ is irrelevant, but the correlations exhibit the fundamental property
\begin{equation}
\begin{split}
 \langle & \psi^{\dag}_{g}({\mathbf r}) \psi_{g}({\mathbf r})   \psi^{\dag}_{g}({\mathbf r'}) \psi_{g}({\mathbf r'}) \rangle =\\
&g^{(2)}({\mathbf r}, {\mathbf r'}) n({\mathbf r}) n({\mathbf r'})+   n({\mathbf r}) \delta({\mathbf r} - {\mathbf r'})
\end{split}
\end{equation}
with $g^{(2)}({\mathbf r}, {\mathbf r'})$ the two-body correlation function. In the present system, the atoms are randomly distributed within the trap and no correlations appear on length scales comparable to
the wave length $\lambda$, i.e., $g_{2} = 1$. Next, we introduce the operators $S^{+}({\mathbf r}) = \psi^{\dag}_{r}({\mathbf r}) \psi_{g}({\mathbf r}) $ creating a Rydberg excitation
from the ground state and  $S^{-}({\mathbf r}) = \psi^{\dag}_{g}({\mathbf r}) \psi_{r}({\mathbf r}) $ for a transition from the Rydberg state into the ground state.
These operators satisfy the commutation relation
\begin{equation}
\begin{split}
&[S^{-}({\mathbf r}) ,S^{+}({\mathbf r'}) ] = \\
&\psi^{\dag}_{g}({\mathbf r}) \psi_{g}({\mathbf r}) \delta({\mathbf r} - {\mathbf r'})- \psi^{\dag}_{r}({\mathbf r}) \psi_{r}({\mathbf r})  \delta({\mathbf r} - {\mathbf r'}).
\end{split}
\end{equation}
Then, the Hamiltonian describing the coupling between light and atomic ensemble within the rotating frame and using the rotating wave approximation
reduces to
\begin{equation}
  H= \int \frac{d{\mathbf q}}{(2 \pi)^3} \hbar \omega_{{\mathbf q}} a^{\dag}_{\mathbf q} a^{}_{\mathbf q}
  + g \int d{\mathbf r} \left[ S^{-}({\mathbf r}) \mathcal{ E}^{\dag}({\mathbf r}) + \mathcal{E}^{}({\mathbf r}) S^{+}({\mathbf r})  \right]   \label{3DHamiltonian}
\end{equation}
with $g$ the dipole matrix element for the optical transition. In the following, we assume a polarization ${\mathbf p}$ of dipole transition  along the x-direction .
Therefore, the electric field operator in three-dimensions takes the form
\begin{equation}
 \mathcal{ E}({\mathbf r}) =\sum_{\mu} \int \frac{d{\mathbf q}}{(2 \pi)^3} \: c^{\mu}_{\mathbf q}\:  a_{\mathbf q} \:  e^{i {\mathbf q} {\mathbf r}}
\end{equation}
with $c^{\mu}_{\mathbf q} = i \sqrt{\omega_{\mathbf q}  2 \pi \hbar  }\:  {\mathbf p} \cdot \varepsilon^{\mu}_{\mathbf q} $ the normalization and influence of the polarization $\varepsilon^{\mu}_{\mathbf q}$.
The incoming electric field is characterized by a Gaussian beam propagating along the z-direction with width $w_{0}$ and polarization parallel to ${\mathbf p}$;
the precise mode function is denoted as $u({\mathbf r})$ and gives rise to the transverse mode area $A= \pi w_{0}^2/2$.
Therefore, this incoming state couples coherently to the $W$-state of the superatom
\begin{equation}
  |W\rangle = \frac{1}{\sqrt{N}} \int d{\mathbf r} \:  u({\mathbf r} )\: S^{+}({\mathbf r} ) |0\rangle .
\end{equation}
Here, $N$ denotes the relevant number of particles overlapping with the incoming mode of the photonic state, i.e.,
\begin{equation}
  N= \int d{\mathbf r} |u({\mathbf r})|^2 \langle 0| \psi^{\dag}_{g}({\mathbf r}) \psi_{g}({\mathbf r}) |0\rangle.
\end{equation}
In general, this quantity varies within each shot of the experiment, as the position of atoms is randomly distributed, but its fluctuations are suppressed by $\Delta N/\bar{N} \sim 1/\sqrt{\bar{N}}$, and can be safely ignored for $10^4$ particles participating in  the superatom;  here $\bar{N}$ is the mean contributing atom number after averaging over many realizations. In the experimentally relevant regime with $\lambda \ll  w_{0},\sigma_{z} $ and $w_{0}\ll \sigma_{r}$, we obtain
\begin{equation}
   \bar{N} = \int d{\mathbf r} |u({\mathbf r})|^2 n({\mathbf r})   = \frac{(2 \pi)^{3/2}}{4} \: w_0^2 \sigma_z n_{0}  = (2 \pi)^{1/2} \sigma_{z} A\:  n_{0} .
   \label{meannumber}
\end{equation}
In order to understand the collective enhancement of the decay of the superatom state $|W\rangle$, we determine its decay rate
within Fermi's Golden rule. The averaged deacy rate into a photonic mode $\mathbf{q}$ with polarization $\varepsilon^{\mu}_{\mathbf q}$
takes the form
\begin{widetext}
\begin{align}
\begin{split}
\bar{\Gamma}_{\mathbf{q},\mu}&= \frac{2 \pi g^2}{ \hbar} \delta(\hbar \omega \! - \! \hbar c |{\mathbf q}|) \: |c^{\mu}_{\mathbf q}|^2 \int d{\mathbf r'}d{\mathbf r} e^{i {\mathbf q} ({\mathbf r}-{\mathbf r'})}
 \left \langle  \frac{\psi^{\dag}_{g}({\mathbf r}) \psi_{g}({\mathbf r})   \psi^{\dag}_{g}({\mathbf r'}) \psi_{g}({\mathbf r'}) }{N}\right \rangle   u^{*}({\mathbf r'})  u({\mathbf r})  \\
 &=    \frac{2 \pi g^2}{ \hbar} \delta(\hbar \omega \!-\! \hbar c |{\mathbf q}|)\:  |c_{\mathbf q}^{\mu}|^2  \left[1 + \frac{1}{\bar{N}} \left|  \int d{\mathbf r} e^{- i {\mathbf q} {\mathbf r}} u({\mathbf r}) n({\mathbf r}) \right|^2  + O(\Delta N/\bar{N})\right] .
\end{split}
\end{align}
\end{widetext}
Here,  $\omega$ denotes the opitcal frequency of the transition.
The first term gives rise to the standard spontaneous decay rate $\Gamma = 4 g^2 \omega^3 / 3\hbar c^3$ for a single atom.
We therefore conclude, that the superatom exhibits an incoherent decay process into an arbitrary photon mode ${\mathbf q}$ giving
rise to the conventional spontaneous decay rate. In turn, the second term  characterizes the  possibility for collective enhancement of the decay into a specific mode.
However, in the experimental parameter regime with $w(\sigma_z) < \sigma_{r}$ this collective decay only provides a significant contribution into the forward direction with an opening angle
\begin{equation}
\sin^2 \theta \lesssim   \frac{1}{ \pi^2} \frac{\lambda^2}{ w_{0}^2}.
\end{equation}
These directions are however comparable to the angular spread of the Gaussian incoming beam. Especially, also the back scattering is suppressed by $\exp(- 8 \pi^2 \sigma_z^2/\lambda^2)$.
Therefore, it is convenient to determine the spontaneous emission of the superatom state into the forward propagating Gaussian beam with mode $u({\mathbf r})$ and polarization $\varepsilon^{\mu}= {\mathbf p}$, i.e.,
\begin{widetext}
\begin{equation}
\kappa = \frac{2 \pi g^2}{\hbar}  \int \frac{d k}{2 \pi} \delta(\hbar \omega - \hbar c k)  \frac{|c^{\mu}_{k}|^2}{ A} \left[1 +\frac{ 1}{\bar{N}} \left|  \int d{\mathbf r} |u({\mathbf r})|^2 n({\mathbf r}) \right|^2 \right] = \frac{2 \pi (\bar{N}+1) g^2}{A} \frac{\omega}{ \hbar c}.
\end{equation}
\end{widetext}
Here, $A = \pi w_{0}^2/2$ denotes the transverse mode volume of the Gaussian beam and in the following discussion we can well approximate $\bar N +1 \approx \bar N$. For a transverse width of the atomic density distribution
$\sigma_{r} \lesssim w_{0}$, transitions into higher Gaussian modes are possible and the determination of the decay rate into these modes is straightforward;
these terms describe the fact that a narrow atomic media in free space acts as a lens for the incoming photons.

We conclude from this analysis, that the superatom state $|W\rangle$ collectively couples to a one-dimensional channel of forward propagating modes. The latter are described by the Gaussian beam of the incoming probe field, and the collective coupling strength takes the form
\begin{equation}
 g_{\rs{ col}}= 2 \sqrt{\kappa} = \sqrt{\frac{8 \pi \bar{N} g^2 \omega}{ A \hbar c}} = \sqrt{\frac{ 3 \bar{N} \: \Gamma \: \lambda^2}{2 \pi A}} .
\end{equation}
The system therefore reduces to the Hamiltonian in \Eref{eq:hamiltonian}, and the expression for the electric field operator in \Eref{eq:el_field}. Note, that in addition the superatom also exhibits an internal dynamics by the virtual exchange of photons \cite{Lehmberg1970,Scully2008}, coupling the $\ket{W}$ state to the dark states. For the narrow bandwidth pulses used in the experiment, the variation in momentum of the electric field is small and we can safely ignore any changes in the transverse wave function of the Gaussian beam. Furthermore, the incoherent  spontaneous emission into the transverse channels as well as back scattering is well accounted for by the single atom decay rate $\Gamma$.

In order to compare the two-level model presented in this appendix with the experimentally obtained results we must consider the adiabatic elimination of the intermediate state present in the experiment. In this case, the effective coupling strength is $g^{\rs eff}_{\rs col} = g_{\rs col} \Omega / 2 \Delta$ and, respectively, $\kappa^{\rs eff} = (g^{\rs eff}_{\rs col})^2/4 = \SI{0.27}{\per\micro\second}$.


\bibliography{bibliography}

\end{document}